\documentclass{elsart}
\usepackage{amsmath, amsfonts, amssymb}
\usepackage{makeidx}

\newtheorem{lemma}[thm]{Lemma}
\newtheorem{corollary}[thm]{Corollary}
\newtheorem{theorem}[thm]{Theorem}

\begin{document}
\begin{frontmatter}

\title{Proof of a Conjecture on the Sequence of Exceptional Numbers, Classifying
Cyclic Codes and APN Functions}
%\author{Fernando Hernando and Gary McGuire}

\author{Fernando Hernando\thanksref{label1}}
\ead{f.hernando@ucc.ie}  
\address{Department of Mathematics\\ 
University College Cork\\
 Ireland}

\author{Gary McGuire\thanksref{label2}}
\ead{gary.mcguire@ucd.ie}
\address{School of Mathematical Sciences\\
University College Dublin\\
Ireland}
\thanks[label1]{Postdoctoral Researcher at
the Claude Shannon Institute, Science Foundation Ireland Grant 06/MI/006,
and also supported by  MEC MTM2007-64704 (Spain).}
\thanks[label2]{Research supported by the Claude
Shannon Institute, Science
Foundation Ireland Grant 06/MI/006.}

\maketitle

\begin{abstract}
We prove a conjecture that classifies exceptional numbers.
This conjecture arises in two different ways,
from cryptography and from coding theory.
An odd integer $t\geq 3$ is said to be exceptional if $f(x)=x^t$ is
APN (Almost Perfect Nonlinear) over $\mathbb{F}_{2^n}$ for
infinitely many values of $n$.
Equivalently, $t$ is exceptional if the binary cyclic code of length $2^n-1$
with two zeros $\omega, \omega^t$ has minimum distance $5$ for
infinitely many values of $n$.
The conjecture we prove states that every exceptional number has
the form $2^i+1$ or $4^i-2^i+1$.
\end{abstract}

\bigskip

\begin{keyword}
Absolutely irreducible polynomial\sep  coding theory\sep  cryptography.
\end{keyword}

\end{frontmatter}

\newpage

\section{Introduction}
 The sequence of numbers of the form $2^i+1$ or $4^i-2^i+1$ (where $i\geq 1$) is
\[
3,5,9,13,17,33,57,65,129, 241, 257, 513, 993, 1025, \ldots .
\]
This  is sequence number A064386 in the On-Line Encyclopedia
of Integer Sequences.
It has been known for almost 40 years that these numbers are \emph{exceptional} numbers,
in the sense we will define shortly.
No further exceptional numbers were found, and
it was conjectured that this sequence  is the complete
list of exceptional numbers.
In this article we complete the proof of this conjecture.
Somewhat surprisingly, the sequence of exceptional numbers  arises in two different contexts,
as explained in the excellent survey article of Dillon \cite{Dillon}.
We now proceed to give these two different motivations
for the conjecture.
\bigskip

\subsection{Coding theory}
%\bigskip

We fix our base field $\mathbb{F}_2$.
Let $w$ be a primitive $(2^n-1)$-th root of unity in an extension of $\mathbb{F}_2$,
i.e., a primitive element of $\mathbb{F}_{2^n}$.
For every odd $t\geq 3$, we
define $C_n^t$ as the cyclic code over $\mathbb{F}_2$ of length
$2^n-1$ with two zeros $w, w^t$.
It is well known that if $t=3$, the code
$C_n^3$ has minimum distance $5$ for every $n\geq 3$.
This code is called the 2-error-correcting BCH code.
We want to find other
 values of $t$ (fixed with respect to $n$)
 for which the code $C_n^t$ has minimum distance
$5$ for infinitely many values of $n$. Those values of
$t$ having this property are called
\textbf{exceptional}.
The only known exceptional values for $t$
are numbers of the form $t=2^i+1$ (known in the field of coding theory as Gold numbers)
and $t=4^i-2^i+1$ (known as Kasami-Welch numbers).
We give more on the precise history in Section \ref{CodingBack}.
The conjecture stated by Janwa-McGuire-Wilson
\cite{Janwa-McGuire-Wilson} is

\bigskip
\textbf{Conjecture 1:} {\em The only exceptional values for $t$ are the
Gold and Kasami-Welch numbers.}

\vspace{.2cm}

Equivalently, the conjecture says that for a fixed odd $t\geq 3$, $t\neq2^i+1$ or $t\neq
4^{i}-2^i+1$, the codes $C_n^t$ of length $2^n-1$ have codewords
of weight $4$ for all but for finitely many values of $n$.
In this paper we prove Conjecture 1.

\bigskip

\subsection{Cryptography}

%\bigskip

The second approach to this problem comes from  cryptography.
One of the desired properties for an S-box used in a block cipher is to have the best
possible resistance against differential attacks, i.e., any given
plaintext difference $a=y-x$ provides a ciphertext difference $f(y)-f(x)=b$
with small probability. More formally, a function
$f:\mathbb{F}_{2^n}\rightarrow \mathbb{F}_{2^n}$ is said to be APN
(Almost Perfect Nonlinear) if for every $a,b\in\mathbb{F}_{2^n}$
with $a\neq 0$ we have
$$
\sharp\{x\in\mathbb{F}_{2^n}\mid f(x+a)+f(x)=b\}\leq 2.
$$
Over a field of characteristic $2$, APN functions provide optimal resistance to
differential cryptanalysis.

Monomial functions $f(x)=x^t$ from
$\mathbb{F}_{2^n}\longrightarrow \mathbb{F}_{2^n}$
are often considered for use in applications.
The exponent $t$ is called \textbf{exceptional} if $f(x)=x^t$ is APN
on infinitely many extension fields of $\mathbb{F}_{2}$.
The conjecture stated by Dillon \cite{Dillon} is

\bigskip
\textbf{Conjecture 2:} {\em The only exceptional exponents are the
Gold and Kasami-Welch numbers.}

\bigskip

Of course, Dillon knew that Conjecture 2 is the same as Conjecture 1, as we explain below.
Conjecture 2 says that for a fixed odd $t\geq 3$, $t\neq2^i+1$ or $t\neq
4^{i}-2^i+1$, the function $f(x)=x^t$ is APN on at most a finite number of
fields $\mathbb{F}_{2^n}$.
In this paper we prove Conjecture 2.

\subsection{Summary of Paper}

In Section 2 we will explain why Conjecture 1 and Conjecture 2 are the same.
Section 3 gives some known results and some background theory
that we will need.
The proof naturally splits into two cases.
We use the notation $t=2^i{\ell}+1$, where $i\geq 1$, and $\ell\geq 3$ is odd.
The two cases are dependent on the value of $gcd(\ell, 2^i-1)$.
In Section 4 we recall a result of Jedlicka, which proves the result in
the case  $gcd(\ell, 2^i-1)<\ell$.
In Section 5 we give a proof of the main theorem of this paper, Theorem \ref{MainTheo},
which proves the case  $gcd(\ell, 2^i-1)=\ell$.
In Section 6 we give a counterexample to Conjecture 3 (stated below).

\section{Background}

 This proofs in this paper concern the absolute irreducibility of certain polynomials.
 In this section we will outline how these polynomials arise
 from Conjectures 1 and 2.

\subsection{Coding Theory}\label{CodingBack}

It is well known that
codewords of weight 4 in $C_n^t$ are equivalent to the polynomial
\begin{equation}\label{PolyF}
f_t(x,y,z)=x^t+y^t+z^t+(x+y+z)^t
\end{equation}
having a rational point $(\alpha,\beta,\gamma)$ over $\mathbb{F}_{2^n}$ with distinct coordinates.
Notice that $x+y$, $x+z$ and $y+z$ divide
$f_t(x,y,z)$, so we may restrict ourselves
to rational points of the homogeneous polynomial
\begin{equation}\label{PolyG}
g_t(x,y,z)=\frac{f_t(x,y,z)}{(x+y)(x+z)(y+z)}.
\end{equation}
Janwa-Wilson \cite{Janwa-Wilson} provide the following result using the Weil bound.
\begin{prop}\label{absirredprop}
If $g_t(x,y,z)$ has an absolutely irreducible factor defined over  $\mathbb{F}_2$
then $g_t(x,y,z)$ has
rational points  $(\alpha,\beta,\gamma)\in (\mathbb{F}_{2^n})^3$ with
distinct coordinates for all $n$ sufficiently large.
\end{prop}
The following conjecture was proposed by Janwa-McGuire-Wilson.
\bigskip

\noindent\textbf{Conjecture 3:} {\em The polynomial $g_t(x,y,z)$ is
absolutely irreducible for all $t$ not of the form $2^i+1$ or
$4^i-2^i+1$.}

\bigskip

We give a counterexample (found with MAGMA) to Conjecture 3 in Section 6.
A slightly weaker form of Conjecture 3 is:
\bigskip

\noindent\textbf{Conjecture 3$'$:} {\em The polynomial $g_t(x,y,z)$ has an
absolutely irreducible factor defined over $\mathbb{F}_2$
for all $t$ not of the form $2^i+1$ or
$4^i-2^i+1$.}

\bigskip

By Proposition \ref{absirredprop} and the discussion above, it is clear that
Conjecture 3 $\Rightarrow$ Conjecture 3$'$ $\Rightarrow$ Conjecture 1.
In this paper we will prove Conjecture 3$'$, and as a result, we prove
Conjecture 1.

Notice that $g_t(x,y,z)$ has no singular points at the infinity, thus the latter conjecture may be reformulated using $g_t(x,y,1)$ instead of $g_t(x,y,z)$.
We write $f_t(x,y)$ for $f_t(x,y,1)$, and we write
$g_t(x,y)$ for $g_t(x,y,1)$.

For the known exceptional values of $t$, that is, when $t$ has the form
$2^i+1$ or $4^i-2^i+1$, the polynomial $g_t(x,y)$ is known to \emph{not}
be absolutely irreducible, and the factorization is described in \cite{Janwa-Wilson}.
We also remark that for some values of $t$, such as $t=7$,
$g_t(x,y)$ is nonsingular and therefore absolutely irreducible,
but it is false that $g_t(x,y)$ is nonsingular for all $t$ not of
the form $2^i+1$ or $4^i-2^i+1$.

\subsection{Cryptography}

It is well known that
$$
h_t(x,y)=\frac{(x+1)^t+x^t+(y+1)^t+y^t}{(x+y)(x+y+1)}.
$$
has no rational points over $\mathbb{F}_{2^n}$ besides those with $x=y$ and $x=y+1$
if and only if $x^t$ is APN over $\mathbb{F}_{2^n}$.

Analogous to Proposition \ref{absirredprop}, Jedlicka \cite{Jedlicka} showed that
as a consequence of the Weil bound we have the following result.
\begin{prop}\label{absirredprop2}
If $h_t(x,y)$ has an absolutely irreducible factor over $\mathbb{F}_2$
then  $h_t(x,y)$
has  rational points over $\mathbb{F}_{2^n}$ besides those with $x=y$ and $x=y+1$
for all $n$ sufficiently large.
\end{prop}

The following conjectures are essentially stated in \cite{Jedlicka}.
\bigskip

\noindent\textbf{Conjecture 4:} {\em The polynomial $h_t(x,y)$ is  absolutely
irreducible polynomial  for all $t$ not of the form $2^i+1$ or
$4^i-2^i+1$.}

\bigskip

A slightly weaker version of this conjecture is:
\bigskip

\noindent\textbf{Conjecture 4$'$:}
{\em The polynomial $h_t(x,y)$ has an
absolutely irreducible factor defined over $\mathbb{F}_2$
for all $t$ not of the form $2^i+1$ or
$4^i-2^i+1$.}
\bigskip

By Proposition \ref{absirredprop2} and the discussion above, it is clear that
Conjecture 4 $\Rightarrow$ Conjecture 4$'$ $\Rightarrow$ Conjecture 2.
In this paper we will prove Conjecture 4$'$, and as a result, we prove
Conjecture 2.  We give a counterexample to Conjecture 4.

\subsection{Putting the Problems Together}

The following is well known to researchers in the area.

\begin{lemma}
Conjecture $3$ is true iff Conjecture $4$ is true.
Conjecture $3 '$ is true iff Conjecture $4'$ is true.
\end{lemma}

Proof: Factoring out $y^t$ from  $(x+1)^t+x^t+(y+1)^t+y^t$ and letting
 $X=\frac{x+1}{y}$ and
$Y=\frac{x}{y}$ gives
$$
(x+1)^t+x^t+(y+1)^t+y^t=y^t[X^t+Y^t+1+(X+Y+1)^t].
$$
Therefore, we can study the irreducibility of $h_t(x,y)$
or that of $g_t(x,y)$, they are equivalent.
$\square$\bigskip

We can say even more: the monomial $x^t$ is  an APN
function over $\mathbb{F}_{2^{n}}$ if and only if  the code
$C_n^t$ has minimum distance $5$.
This shows that Conjecture 1 is true iff Conjecture 2 is true.
\bigskip

\begin{tabular}{ccc}
Conjecture 3 & $\iff$ & Conjecture 4\\
$\Downarrow$ & & $\Downarrow$\\
Conjecture $3'$ & $\iff$ & Conjecture $4'$\\
$\Downarrow$ & & $\Downarrow$\\
Conjecture $1$ & $\iff$ & Conjecture $2$\\
\end{tabular}

\bigskip
In this paper we will prove Conjecture 3$'$,  see the table below.
This is equivalent to proving Conjecture 4$'$, and so
implies both Conjectures 1 and 2.
However, we give a counterexample in Section 6
that shows that Conjectures 3 and 4 are false in general.

 \bigskip

\textbf{Notation:} Throughout we will let
$t=2^i{\ell}+1$, where $i\geq 1$, and $\ell\geq 3$ is odd.

We use the notation $Sing(g_t)$ to denote the set of all singular points of $g_t$.

\bigskip

The following box summarizes known results before this paper,
and what is done in this paper.
\begin{center}
\begin{tabular}{|l | c| c | c|}
\hline

Function & Exceptional & Constraints & Author \\
\hline\hline
$x^{2^i+1}$ & Yes & APN iff $(i,n)=1$ & Gold \cite{Gold}\\
\hline
$x^{4^{i}-2^i+1}$ & Yes & APN iff $(i,n)=1$ & van Lint-Wilson \cite{van Lint-Wilson}, \\
&&&Janwa-Wilson \cite{Janwa-Wilson}\\
\hline
$x^{t}$ & No & $t\equiv 3(mod \ 4), t>3$ & Janwa-McGuire-Wilson \cite{Janwa-McGuire-Wilson}\\
\hline $x^{2^i{\ell}+1}$ & No & $gcd(\ell , 2^i-1)<\ell$ & Jedlicka \cite{Jedlicka} \\
&&&\\
\hline
$x^{2^i\ell+1}$ & No & $gcd(\ell, 2^i-1)=\ell$ & This paper \\ \hline
\end{tabular}
\end{center}
\bigskip

Janwa-McGuire-Wilson proved the $i=1$ case.
The full proof of Conjecture 3$'$ divides into two cases, according as
$gcd(\ell , 2^i-1)<\ell$  or $gcd(\ell , 2^i-1)=\ell$.
Jedlicka  proved the case $gcd(\ell , 2^i-1)<\ell$.
In the present work we give a proof of Conjecture 3$'$ in the remaining case 
when $gcd(\ell , 2^i-1)=\ell$.
We show that Conjecture 3 is false in general. This completes  the
classification of exceptional exponents.

%%%%%%%%%%%%%%%%%%%%%%%%%%%%%%%%%%%%%%%%%%%%%%%%%%%%%%%%%%%%%%%%%%%%%%%%%%%%%%%%%%%%%%%%%%%%%%%%%%%%%%%%%%%%%%%%%
%%%%%%%%%%%%%%%%%%%%%%% END OF SECTION : Introduction   %%%%%%%%%%%%%%%%%%%%%%%%%%%%%%%%%%%%%%%%%%%%%%%%%%%%%%%
%%%%%%%%%%%%%%%%%%%%%%%%%%%%%%%%%%%%%%%%%%%%%%%%%%%%%%%%%%%%%%%%%%%%%%%%%%%%%%%%%%%%%%%%%%%%%%%%%%%%%%%%%%%%%%%%%

\section{Singularities and Bezout's Theorem}
Consider $P=(\alpha,\beta)$, a point in the plane.
Write
$$
f_t(x+\alpha,y+\beta)=F_0+F_1+F_{2}+F_{3}+\cdots
$$
where $F_m$ is homogeneous of degree $m$.
The multiplicity of $f_t$ at $P$ is the smallest $m$ with $F_m\not=0$,
and is denoted by $m_P(f_t)$.
In this case, $F_m$ is called the tangent cone.

Recall the notation that
$t=2^i{\ell}+1$, where $i\geq 1$, and $\ell\geq 3$ is odd.

We let
$\lambda=\alpha+\beta+1$, then straightforward calculations
\cite{Janwa-Wilson} give
$$
F_0=\alpha^t+\beta^t+\lambda^t+1
$$
$$
F_1=(\alpha^{t-1}+\lambda^{t-1})x+(\beta^{t-1}+\lambda^{t-1})y
$$
$$
F_{2^i}=(\alpha^{t-2^i}+\lambda^{t-2^i})x^{2^i}+(\beta^{t-2^i}+\lambda^{t-2^i})y^{2^i}
$$
$$
F_{2^i+1}=(\alpha^{t-2^i-1}+\lambda^{t-2^i-1})x^{2^i+1}+(\beta^{t-2^i-1}+\lambda^{t-2^i-1})y^{2^i+1}+\lambda^{t-2^i-1}(x^{2^i}y+xy^{2^i})
$$
and $F_j=0$ for $1<j<2^i$.
A point $P=(\alpha,\beta)$ is singular if and only if $F_0=F_1=0$,
which happens if and only if
$\alpha,\beta$ and $\lambda$ are $\ell$-th roots of unity (see \cite{Janwa-Wilson}).
We distinguish  three types of singular point.
\begin{enumerate}
\item[(I)] $\alpha=\beta=\lambda=1$.

\item[(II)] Either $\alpha=1$ and $\beta\neq 1$, or
$\beta=1$ and $\alpha\neq 1$, or
$\alpha=\beta\neq 1$ and
$\lambda=1$.

We divide these singular points into two cases:

(II.A) Where II holds and $\alpha, \beta \in GF(2^i)$

(II.B) Where II holds and $\alpha, \beta$ not both in  $GF(2^i)$.

\item[(III)] $\alpha\neq 1$, $\beta\neq 1$ and $\alpha\neq \beta$.

We divide these singular points into two cases:

(III.A) Where III holds and $\alpha, \beta \in GF(2^i)$

(III.B) Where III holds and $\alpha, \beta$ not both in  $GF(2^i)$.

\end{enumerate}

Now we summarize some properties already known, for more details
see \cite{Janwa-McGuire-Wilson}.

\begin{lemma}\label{LinearFactors}
If $F_{2^i}\neq 0$ then $F_{2^i}=(Ax+By)^{2^i}$ where
$A^{2^i}=\alpha^{1-2^i}+\lambda^{1-2^i}$ and
$B^{2^i}=\beta^{1-2^i}+\lambda^{1-2^i}$.
\end{lemma}

The proof is obvious, because we are in characteristic $2$.
The importance of this lemma is that there is only one distinct
linear factor in $F_{2^i}$.
Another useful fact is that the opposite is true for $F_{2^{i}+1}$,
as shown in \cite{Janwa-McGuire-Wilson}:

\begin{lemma}\label{DifferentLinearFactors}
$F_{2^{i}+1}$ has $2^i+1$ distinct linear factors.
\end{lemma}

\subsection{Classification of Singularities}

The next step is to describe how many
singularities of each type there are, and to find their
multiplicities.

Clearly there is only one singularity of type I. There are
$(\ell-1)$ points of type $(1,\beta)$ with $\beta^{\ell}=1$ and
$\beta\neq 1$. So, there are also $(\ell-1)$ of type $(\alpha,1)$
and $(\ell-1)$ of type $(\alpha,\alpha)$ with $\alpha^{\ell}=1$
and $\alpha\neq 1$. In total there are $3(\ell-1)$ points of type
II.

For points of type III there are $(\ell-1)$ choices for
$\alpha\neq 1$, and thus there are $(\ell-2)$ choices for $\beta$
with $\beta\neq 1$ and $\beta\neq \alpha$.
However, not all these choices lead to a valid singular point.
We upper bound the number of valid choices in the next lemma.

\begin{lemma}
For every $\alpha$ with $\alpha^{\ell}=1$ and $\alpha\neq 1$ there
exists a $\beta$ with $\beta^{\ell}=1$, $\beta\neq \alpha$ and $\beta\neq 1$
such that $(\alpha+\beta+1)^{\ell}\neq 1$.
\end{lemma}
\textbf{Proof:}  Suppose the statement is false, and
fix an $\alpha\not=1$ such that for all $\beta$ with $\beta^{\ell} = 1$
we also have  $(\alpha+\beta+1)^{\ell}= 1$.
Let $H$ be $\{a \mid a^{\ell}=1\}$, the set of ${\ell}$-th roots of
unity. Consider the map,
\begin{equation*}
\phi:H\rightarrow H, \ \phi(\beta)=\alpha+\beta+1.
\end{equation*}
The key point is that this map has no fixed points. For, if
$\phi(\beta)=\beta$, then $\alpha=1$, which is not true by assumption. Thus $\phi$  is a
permutation of $H$ which is a product of transpositions of the form
$(\beta, 1+\alpha+\beta)$. Therefore $\phi$ must
permute an even number of points, which contradicts the fact that
${\ell}$ is odd. $\square$

\bigskip

From this lemma if follows that, given $\alpha$,
there are at most $({\ell}-3)$ possible
choices for $\beta$. We can not guarantee that
each of these is valid, so we can only upper bound the points of type III by
$\leq ({\ell}-1)({\ell}-3)$.
There are cases when this bound is tight.

The next Lemma helps us determine when $m_P (f_t)$ is equal to $2^i$ and
when it is $2^i+1$.

\begin{lemma}\label{FirstCoef}
Let $P=(\alpha,\beta)$ be a singular point of $f_t$, then
$F_{2^i}=0$ if and only if one of the following holds.
\begin{enumerate}
\item $P$ is of Type I
\item $P$ is of Type II.A
\item $P$ is of Type III.A
\item $P$ is of Type III.B and $\alpha/\beta$ and $\beta/\lambda \in GF(2^i)$.
In this case, we have $1<gcd(\ell, 2^i-1)<\ell$.
\end{enumerate}
\end{lemma}

\textbf{Proof:} We have to check when
$\alpha^{t-2^i}+\lambda^{t-2^i}=0$. Substituting $t=2^i{\ell}+1$
in the formula we get  $\alpha^{1-2^i}=\lambda^{1-2^i}$, or
$\alpha^{2^i-1}=\lambda^{2^i-1}$. Now reasoning with
$\beta^{t-2^i}+\lambda^{t-2^i}=0$ we also obtain that either
$\beta^{2^i-1}=\lambda^{2^i-1}$.
So $F_{2^i}=0$ if and only if
$\alpha^{2^i-1}=\beta^{2^i-1}=\lambda^{2^i-1}$.
Consequently,  $F_{2^i}=0$ if and only if
$(\alpha/\beta)^{2^i-1}=(\beta/\lambda)^{2^i-1}$.

If $P$ is of Type I or II.A or III.A, then in fact
$\alpha^{2^i-1}=\beta^{2^i-1}=\lambda^{2^i-1}=1$.
If $P$ is of Type II.B then $F_{2^i}\not=0$ because certainly
one coefficient does not vanish.
Finally, suppose $P$ is of Type III.B, and then
we may deduce that $\alpha=C\beta$ and $\beta=D\lambda$
for some $C,D \in GF(2^i)$.
Raising to the $\ell$-th power
yields that $C,D$ are $\ell$-th roots of unity.
Letting $d=gcd(\ell, 2^i-1)$, then $C,D$ are $d$-th roots of unity.
Because $C$ and $D$ cannot be 1, we must have $d>1$.
If $d=\ell$ then all $\ell$-th roots of unity are in $GF(2^i)$.
Because $P$ is of Type III.B, at least one of
$\alpha, \beta$ is not in $GF(2^i)$,
so $d<\ell$.
 $\square$
\bigskip

Note that if $\ell=2^i-1$ then $t=2^i \ell+1=4^i-2^i+1$, which is an
exceptional value.

We now list the classification in a table.
We let $w(x,y)=(x+1)(y+1)(x+y)$ so that $f_t=wg_t$ and
$m_P(f_t)=m_P(g_t)+m_P(w)$.
The values of $m_P(w)$ are easy to work out for the various singular points $P$.
The implications of Lemma \ref{FirstCoef} can be summarized in the following tables.

\begin{itemize}
\item[$gcd(\ell, 2^i-1)=1$]
\begin{center}
\begin{tabular}{|l | c| c | c|}
\hline
Type & Number of Points & $m_P(f_t)$ & $m_P(g_t)$ \\
\hline
I & $1$ & $2^{i}+1$ & $2^{i}-2$\\
\hline
II & $3({\ell}-1)$ & $2^i$ & $2^i-1$\\
\hline
III & $\leq ({\ell}-1)({\ell}-3)$ & $2^i$ & $2^i$\\
\hline
\end{tabular}
\end{center}
In this case, the Type II points are all of Type II.B, and the Type III points are all of Type III.B.
\bigskip
\item[$gcd(\ell, 2^i-1)=\ell$]
\begin{center}
\begin{tabular}{|l | c| c | c|}
\hline
Type & Number of Points & $m_P(f_t)$ & $m_P(g_t)$ \\
\hline
I & $1$ & $2^{i}+1$ & $2^{i}-2$\\
\hline
II & $3({\ell}-1)$ & $2^i+1$ & $2^i$\\
\hline
III & $\leq ({\ell}-1)({\ell}-3)$ & $2^i+1$ & $2^i+1$\\
\hline
\end{tabular}
\end{center}
In this case, the Type II points are all of Type II.A, and the Type III points are all of Type III.A.
\end{itemize}
The case $1<gcd(\ell, 2^i-1) < \ell $
is a mixture of the previous two
cases because $f_t(x,y)$ has points with multiplicity $2^i$ and points with multiplicity $2^{i}+1$.
Nevertheless the upper bounds on the \emph{number} of points still hold.

\subsection{Bezout's Theorem}

One of the central results in our work uses  Bezout's theorem, which is a classical result in algebraic geometry and appears frequently in the literature \cite{Fulton}.

\bigskip

{\textbf{Bezout's Theorem}:} Let $r$ and $s$ be two
projective plane curves over a
field $k$ of degrees $D_1$ and $D_2$ respectively having no components in common. Then,
\begin{equation}\label{Bezout}
\sum_{P}I(P,r,s)=D_1D_2.
\end{equation}
The sum runs over all the points $P=(\alpha,\beta)\in
\overline{k}\times\overline{k}$, and by $I(P,r,s)$ we understand
the intersection multiplicity of the curves $r$ and $s$ at the
point $P$. Notice that if $r$ or
$s$ does not go through $P$, then $I(P,r,s)=0$. Therefore, the sum
in (\ref{Bezout}) runs over the singular points of the product $rs$.

Using properties $I(P,r_1r_2,s)=I(P,r_1,s)+I(P,r_2,s)$ and
$\deg (r_1r_2)=\deg (r_1) +\deg (r_2)$ one can generalize
Bezout's Theorem to several curves $f_1$, $f_2$, $\cdots$ ,$f_r$:
\begin{equation}\label{bezoutseveral}
\sum_P \sum_{1\leq i<j\leq r}  I(P,f_j,f_j)= \sum_{1\leq i<j\leq r }\deg (f_j)\deg (f_j).
\end{equation}

The following property of the intersection multiplicity will be useful for us.
It is part of the definition of intersection multiplicity in \cite{Fulton}.
We state it as a Corollary.

\begin{corollary}\label{DifferentTangentCone}
\begin{equation}\label{IntersectionMultiplicity}
I(P,r,s)\geq m_P(r)m_P(s),
\end{equation}
and  equality holds if and only if the tangent cones of $r$ and $s$ do not share any linear factor.
\end{corollary}

Janwa-McGuire-Wilson \cite{Janwa-McGuire-Wilson} have  computed the
intersection multiplicity at points of type II.B
assuming the curve $g_t(x,y)$ factors:
\begin{lemma}\label{NoIntersectionPointsTypeII}
If $P$ is a point of type II.B  and $g_t(x,y)=r(x,y)s(x,y)$ then $I(P,r,s)=0$.
\end{lemma}

%%%%%%%%%%%%%%%%%%%%%%%%%%%%%%%%%%%%%%%%%%%%%%%%%%%%%%%%%%%%%%%%%%%%%%%%%%%%%%%%%%%%%%%%%%%%%%%%%%%%%%%%%%%%%%%%%
%%%%%%%%%%%%%%%%%%%%%%% END OF SECTION: Singularities   %%%%%%%%%%%%%%%%%%%%%%%%%%%%%%%%%%%%%%%%%%%%%%%%%%%%%%%%%
%%%%%%%%%%%%%%%%%%%%%%%%%%%%%%%%%%%%%%%%%%%%%%%%%%%%%%%%%%%%%%%%%%%%%%%%%%%%%%%%%%%%%%%%%%%%%%%%%%%%%%%%%%%%%%%%%

\section{The Case $gcd(\ell, 2^i-1)<\ell$}

Jedlicka has proved in \cite{Jedlicka} (see Theorem 1) that $g_t(x,y)$ has an absolutely irreducible factor over $\mathbb{F}_2$ whenever  $gcd(\ell, 2^i-1)<\ell$ and $t$ 
is not a Gold or Kasami-Welch number.
Therefore, if we prove the same for the case $gcd(\ell, 2^i-1)=\ell$ then  conjecture $3'$ is 
completely proved.
This is what we do in the next section.

%%%%%%%%%%%%%%%%%%%%%%%%%%%%%%%%%%%%%%%%%%%%%%%%%%%%%%%%%%%%%%%%%%%%%%%%%%%%%%%%%%%%%%%%%%%%%%%%%%%%%%%%%%%%%%%%%
%%%%%%%%%%%%%%%%%%%%%%% END OF SECTION: Case gcd(\ell, 2^i-1)<\ell  %%%%%%%%%%%%%%%%%%%%%%%%%%%%%
%%%%%%%%%%%%%%%%%%%%%%%%%%%%%%%%%%%%%%%%%%%%%%%%%%%%%%%%%%%%%%%%%%%%%%%%%%%%%%%%%%%%%%%%%%%%%%%%%%%%%%%%%%%%%%%%%

\section{Main Result:  Case  $gcd(\ell, 2^i-1)=\ell$}

The principal starting observation when  $gcd(\ell, 2^i-1)=\ell$ is to notice that
all $\ell$-th roots of unity lie in $GF(2^i)$.  
Therefore,  all Type II singularities have Type II.A, and
all Type III singularities have Type III.A.
From Lemma \ref{FirstCoef}, or the table following it, 
we have that $F_{2^i}=0$ at all singular points.

\subsection{Preliminary Lemmata}

One of the main ideas involved in the proof in this section is that if $g_t(x,y)$ is
irreducible over $\mathbb{F}_2$ and splits in several factors
(over an extension field), then  all factors have the same degree.
 The next lemma concerns this sort of phenomenon, and its proof can
be found in \cite{Kopparty-Yekhanin} (although it is surely older).

\begin{lemma}\label{IfIrreducibleEqualDegreeFactors}
Suppose that $p(\underline{x})\in \mathbb{F}_q[x_1,\ldots,x_n]$ is of degree $t$ and is irreducible in $\mathbb{F}_q[x_1,\ldots,x_n]$. There there exists $r\mid t$ and an absolutely irreducible polynomial
$h(\underline{x})\in\mathbb{F}_{q^r}[x_1,\ldots,x_n]$ of degree $\frac{t}{r}$ such that
$$
p(\underline{x})=c\prod_{\sigma\in G}\sigma(h(\underline{x})),
$$
where $G=Gal(\mathbb{F}_{q^r}/\mathbb{F}_{q})$ and $c\in \mathbb{F}_{q}$. Furthermore if $p(\underline{x})$
is homogeneous, then so is $h(\underline{x})$.
\end{lemma}

Here are some more lemmata we will use.

\begin{lemma}\label{MaxOfMany}
Given $N\in \mathbb{N}$ the values  $x_1,\ldots,x_n$ that maximize the function $H(x_1,\ldots,x_n)=\sum_{\substack {1\leq i<j\leq n\\ i\neq j}} x_ix_j$
subject to the constraint
$x_1+\cdots+x_n=N$ are $x_1=\cdots=x_n=N/n$.
\end{lemma}

One more technical result is recorded now, whose proof is trivial.

\begin{lemma}\label{TechnicalResult}
If $i> 2$ and  ${\ell}\mid 2^i-1$ but ${\ell}\neq 2^i-1$ then the following
results hold:
\begin{itemize}
\item[(1)] $2^{i-1}+1-{\ell}>2$. \item[(2)]
$\frac{{\ell}-3}{2^{i+1}}<\frac{1}{4}.$
\end{itemize}
\end{lemma}
\textbf{Proof:} Since  ${\ell}\mid 2^i-1$ but ${\ell}\neq 2^i-1$,
and both numbers are odd,
we certainly have that ${\ell}< 2^{i-1}-1.$ Then
$2^{i-1}-1-{\ell}> 0$ so  $2^{i-1}+1-{\ell}> 2$,
thus (1) holds.

For (2) we have that ${\ell}< 2^{i-1}-1< 2^{i-1}+3$ which implies
$\frac{{\ell}-3}{2^{i-1}}<1$ which certainly implies
$\frac{{\ell}-3}{2^{i+1}}<\frac{1}{4}$.  $\square$

\subsection{A Warm-Up Case}

\begin{theorem}\label{DoNotSplitInTwo}
Suppose that $g_t(x,y)$ is irreducible over $\mathbb{F}_2$ and
${\ell}\mid 2^i-1$ but ${\ell}\neq 2^i-1$. Then $g_t(x,y)$ can not
split in two factors $g_1$ and $g_2$ with $deg(g_1)=deg(g_2)$.
\end{theorem}

\textbf{Proof:}
We apply Bezout's Theorem, which states
\[
\sum_{P \in Sing(g_t)} I(P,g_1,g_2)=deg(g_1)deg(g_2).
\]
By Lemma \ref{FirstCoef} we know that $F_{2^i}=0$.
Since the tangent cones have different lines
by Lemma \ref{LinearFactors}, Corollary
\ref{DifferentTangentCone} tells us that the left hand side is
equal to $\sum_{P \in Sing(g_t)} m_P(g_1)m_P(g_2)$.  Using the
table of singularities described in Section 3 for ${\ell}\mid 2^i-1$ we get
\begin{multline}\label{eq1}
 \sum_{P \in Sing(g_t)} m_P(g_1)m_P(g_2)\leq \\(2^{i-1}-1)^2+3({\ell}-1)2^{2i-2}+({\ell}-1)({\ell}-3)2^{i-1}(2^{i-1}+1).
\end{multline}
 Since the degrees of both components are the same, the right hand side
 of Bezout's Theorem  is exactly,
\begin{equation}\label{eq2}
(2^{i-1}{\ell}-1)^2=2^{2i-2}{\ell}^2-2{\ell}2^{i-1}+1.
\end{equation}
Let us compare $(\ref{eq2})$ and $(\ref{eq1})$. If $(\ref{eq2})>(\ref{eq1})$, we have won,
and this happens if and only if,

\begin{equation}\label{eq3}
2^{2i-2}(-{\ell}+1)+2^{i-1}({\ell}^2-2{\ell}+1)<0
\end{equation}
which is equivalent to
\begin{equation}
2^{i-1}({\ell}-1)>({\ell}^2-2{\ell}+1)=({\ell}-1)^2.
\end{equation}
So we conclude that the condition for $(\ref{eq2})>(\ref{eq1})$ is
\begin{equation}\label{eq4}
2^{i-1}>({\ell}-1)
\end{equation}
which is true by Lemma \ref{TechnicalResult} part (1).  $\square$\bigskip

Remark:
Notice that this proof fails when ${\ell}=2^i-1$, as it should.

The key idea in the previous proof is to compare $(\ref{eq2})$ and
$(\ref{eq1})$. In the next result we have a sharper bound which
will be very useful for further results.

\begin{lemma}\label{SquareOfDegrees&SquareOfMultiplicities}
If ${\ell}\mid 2^i-1$ but ${\ell}\neq 2^i-1$,  then
$$
\deg (g_t)^2> \sum_{P\in Sing(g_t)} m_p(g_t)^2.
$$
\end{lemma}
\textbf{Proof:} Suppose not. Then,
\begin{eqnarray*}
\deg (g_t)^2&=&(2^i\ell-2)^2\\ &\leq& \sum_{P\in Sing(g_t)} m_p(g_t)^2\\
&\leq&
(2^i-2)^2+(3{\ell}-3)2^{2i}+({\ell}-1)({\ell}-3)(2^i+1)^2
\end{eqnarray*}
where the last inequality is obtaining using the table of singularities described in section 3.
After rearrangement we obtain,
$$
0\leq
2^{2i}+{\ell}^22^{i+1}+{\ell}^2-{\ell}2^{2i}-4{\ell}2^i-4{\ell}+2^{i+1}+3.
$$
Equivalently,
$$
0\leq 2^i(2({\ell}-1)^2)-2^i({\ell}-1))+({\ell}-1)({\ell}-3).
$$
Dividing by $({\ell}-1)$ we get
$$
2^{i+1}(2^{i-1}-({\ell}-1))\leq {\ell}-3
$$
or
$$
2^{i-1}-({\ell}+1)\leq \frac{{\ell}-3}{2^{i+1}}.
$$
However, by Lemma \ref{TechnicalResult} we know that the left hand
side is a positive integer and right hand side satisfies  $0<
\frac{{\ell}-3}{2^{i+1}}\leq 1/4$, a contradiction.   $\square$\bigskip

Remark:
Again we note that this proof fails if $\ell = 2^i-1$, as it should.

\subsection{Proof Assuming Irreducibility over $\mathbb{F}_2$}

Next we prove Conjecture $3$ under the assumption in the title.

\begin{theorem}\label{AbsIrreIfIrre}
If $g_t(x,y)$ is irreducible over $\mathbb{F}_2$, and ${\ell}\mid
2^i-1$ but ${\ell}\neq 2^i-1$, then $g_t(x,y)$ is absolutely
irreducible.
\end{theorem}
\textbf{Proof:}
Suppose that $g_t(x,y)$ is irreducible over $\mathbb{F}_2$, and
that $g(x,y)=f_1\cdots f_r$ over some extension field of $\mathbb{F}_2$.
By Lemma \ref{IfIrreducibleEqualDegreeFactors}, each $f_i$ has the same degree,
which must be $\deg (g_t)/r$.
If $r$ is even then by letting $g_1=f_1\cdots f_{r/2}$ and
$g_2=f_{1+r/2}\cdots f_r$ we are done  by Theorem \ref{DoNotSplitInTwo}.
 We may therefore assume that $r$ is odd (although our argument
 does not use this, and is also valid when $r$ is even).

We apply (\ref{bezoutseveral}) obtaining
\begin{equation}\label{eq5}
\sum_P \sum_{1\leq i<j\leq r}  I(P,f_j,f_j)= \sum_{1\leq i<j\leq r } \deg(f_j)\deg (f_j).
\end{equation}
The sum over $P$ is over all singular points of $g_t$.
Since the degree of $f_i$ is equal to the degree of $f_j$ then the right hand side is
\begin{equation}\label{eq6}
\sum_{1\leq i<j\leq r }\deg(f_j)\deg(f_j)=\binom{r}{2}
\biggl( \frac{\deg(g_t)}{r} \biggr)^2=\frac{r-1}{2r}\deg(g_t)^2.
\end{equation}

Now we estimate the inner sum on the left hand side of (\ref{eq5}).
For any $P\in Sing(g_t)$,
since $F_{2^{i}+1}$ consists of $2^i+1$
different lines by Lemma \ref{DifferentLinearFactors},
we have $I(P,f_i,f_j)=m_P(f_i)m_P(f_j)$ for any $i,j$ by Corollary \ref{DifferentTangentCone}.
Therefore
\begin{equation}\label{eq7}
\sum_{1\leq i<j\leq r }I(P,f_j,f_j)=\sum_{1\leq i<j\leq r }m_P(f_j)m_P(f_j).
\end{equation}

We maximize (\ref{eq7}) using Lemma \ref{MaxOfMany}. We obtain the upper bound
\begin{equation}\label{eq8}
\sum_{1\leq i<j\leq r }m_P(f_j)m_P(f_j)\leq \binom{r}{2}\biggl( \frac{m_P(g_t)}{r}\biggr)^2=
\frac{r-1}{2r}m_P(g_t)^2.
\end{equation}
We denote by $I,II$ and $III$ the set of singular points of type I,II and III respectively.
Then left hand side in (\ref{eq5}) is equal to
\begin{multline}\label{eq9}
\sum_{P\in I}\sum_{1\leq i<j\leq r}I(P,f_j,f_j)+ \sum_{P\in II}\sum_{1\leq i<j\leq r}I(P,f_j,f_j)
+\\\sum_{P\in III}\sum_{1\leq i<j\leq r}I(P,f_j,f_j) \overset{(\ref{eq8})}{\leq}\\
\sum_{P\in I}\frac{r-1}{2r}m_P(g)^2+\sum_{P\in II}\frac{r-1}{2r}m_P(g)^2+\sum_{P\in III}\frac{r-1}{2r}m_P(g)^2\leq\\
\frac{r-1}{2r}\biggl((2^i-2)^2+(2^i)^2(3{\ell}-3)+(2^i+1)^2(\ell -1)(\ell -3))\biggr).
\end{multline}
Once again, the last inequality is thanks to the table
with classification of singularities given in
section 3 for ${\ell}\mid 2^i-1$.
If $(\ref{eq6})>(\ref{eq9})$ then we have won.
After canceling the factors of $(r-1)/2r$, the inequality $(\ref{eq6})>(\ref{eq9})$  is
\[
(2^i\ell-2)^2>
(2^i-2)^2+(2^i)^2(3{\ell}-3)+(2^i+1)^2({\ell}^2-4{\ell}+3)
\]
which is true because it is
exactly the same inequality as that in the
proof of Lemma \ref{SquareOfDegrees&SquareOfMultiplicities}.
$\square$

\subsection{Proof of Conjecture $3'$}

In this section we will finally prove Conjecture $3'$.

\begin{theorem}\label{MainTheo}
If ${\ell}\mid 2^i-1$ but ${\ell}\neq 2^i-1$, then $g_t(x,y)$
always has an absolutely irreducible factor over $\mathbb{F}_2$.
\end{theorem}

Proof:
Suppose $g_t=f_1\cdots f_r$ is the factorization into irreducible factors over $\mathbb{F}_2$.
Let $f_k=f_{k,1}\cdots f_{k,n_k}$ be the factorization of $f_k$ into $n_k$ absolutely irreducible factors.
Each $f_{k,j}$ has degree $\deg (f_k)/n_k$, by
Lemma \ref{IfIrreducibleEqualDegreeFactors}.

Let us prove an auxiliary result.
\begin{lemma}\label{Deg<Mult}
All $\mathbb{F}_2$-irreducible components $f_k(x,y)$ of $g_t(x,y)$ satisfy the following conditions:
\begin{itemize}
\item
\begin{equation}\label{eq10}
\deg(f_k)^2\leq \sum_{P \in Sing(g_t)}  m_P(f_k)^2.
\end{equation}
\item
\begin{equation}\label{eq11}
\sum_{1\leq i<j\leq n_k}  m_P(f_{k,i})m_P(f_{k,j})\leq m_P(f_k)^2\frac{n_k-1}{2n_k}.
\end{equation}
\end{itemize}
\end{lemma}
\textbf{Proof:}
Applying Bezout's theorem to $f_k$ gives
\begin{equation}\label{eq12}
\sum_{1\leq i<j\leq n_k} \sum_{P \in Sing(f_k)}
I(P,f_{k,i},f_{k,j})=\sum_{1\leq i<j\leq n_k}   \deg(f_{k,i})
\deg(f_{k,j})=\deg(f_k)^2\frac{n_k-1}{2n_k}.
\end{equation}
Since for every $i,j\in\{1,\ldots,n_k\}$ the tangent cones of $f_{k,i}$ and $f_{k,j}$ consist of
different lines by Lemma \ref{LinearFactors},  the left hand side of (\ref{eq12}) is
\begin{equation}\label{eq13}
\sum_{1\leq i<j\leq n_k} \sum_{P \in Sing(f_k)}
I(P,f_{k,i},f_{k,j})= \sum_{P \in Sing(f_k)} \sum_{1\leq i<j\leq n_k}  m_P(f_{k,i})
m_P(f_{k,j})
\end{equation}
by Corollary
\ref{DifferentTangentCone}.
We fix $P$ a singular point.
Applying Lemma \ref{MaxOfMany} to $$\sum_{1\leq i<j\leq n_k}m_P(f_{k,i}) m_P(f_{k,j})$$ subject to
$\sum_{i=1}^{n_k} m_P(f_{k,i})=m_P(f_k)$ we get that
$$
\sum_{1\leq i<j\leq n_k}  m_P(f_{k,i})m_P(f_{k,j})\leq m_P(f_k)^2\frac{n_k-1}{2n_k}
$$
which proves (\ref{eq11}).
Summing over $P$ then proves (\ref{eq10}).
$\square$\bigskip

\textbf{Proof of Theorem \ref{MainTheo}:}

We apply Bezout's Theorem (equation (\ref{bezoutseveral})) one more time  to
the product
$$f_1f_2\ldots f_r=(f_{1,1}\ldots f_{1,n_1})(f_{2,1}\ldots f_{2,n_2})\ldots (f_{r,1}\ldots f_{r,n_r}).
$$
The sum of the intersection multiplicities
(left hand side of equation (\ref{bezoutseveral}))
can be written
\[
\sum_{k=1}^r \sum_{1\leq i<j\leq n_k} \sum_{P \in Sing(g_t)}  I(P,f_{k,i},f_{k,j})+
\sum_{1\leq k<l\leq r} \sum_{\substack{1\leq i\leq n_k\\1\leq j\leq n_l}}
\sum_{P \in Sing(g_t)}  I(P,f_{k,i},f_{l,j})
\]
where the first term is for factors within each $f_k$, and the second term
is for cross factors between $f_k$ and $f_l$.
Since for every $k$ and $i$
the tangent cones of the $f_{k,i}$ consist of different lines by Lemma \ref{LinearFactors},
the previous sums can be written
\begin{equation}\label{eq17}
\sum_{P \in Sing(g_t)}\biggl[
\sum_{k=1}^r \sum_{1\leq i<j\leq n_k}  m_P(f_{k,i})m_P(f_{k,j})+
\sum_{1\leq k<l\leq r} \sum_{\substack{1\leq i\leq n_k\\1\leq i\leq n_l}}   m_P(f_{k,i})m_P(f_{l,j})\biggr].
\end{equation}

Note that
\begin{eqnarray*}
(m_P (g_t))^2&=&\biggl( \sum_{k=1}^r  m_P (f_k) \biggr)^2\\
&=&  \sum_{k=1}^r m_P (f_k)^2 +2 \biggl( \sum_{1\leq k<l\leq r} m_P (f_k) m_P (f_l) \biggr)\\
&=&   \sum_{k=1}^r m_P (f_k)^2 +2 \sum_{1\leq k<l\leq r} \biggl( \sum_{i=1}^{n_k} m_P (f_{k,i})\biggr)
  \biggl( \sum_{j=1}^{n_l} m_P (f_{l,j})\biggr)\\
  &=&  \sum_{k=1}^r m_P (f_k)^2 +2
  \sum_{1\leq k<l\leq r} \sum_{\substack{1\leq i\leq n_k\\1\leq j\leq n_l}}m_P(f_{k,i})m_P(f_{l,j}).
\end{eqnarray*}
Substituting, (\ref{eq17}) becomes
\begin{equation}\label{eq77}
\sum_{P \in Sing(g_t)}\biggl[
\sum_{k=1}^r \sum_{1\leq i<j\leq n_k}  m_P(f_{k,i})m_P(f_{k,j})+
\frac{1}{2}\biggl( m_P(g_t)^2- \sum_{k=1}^r m_P (f_k)^2 \biggr)\biggr].
\end{equation}
Substituting (\ref{eq11}) this is
\begin{eqnarray}\label{eq18}
&\leq &\sum_{P \in\ Sing(g_t)}\biggl[
\sum_{k=1}^r m_P(f_k)^2\frac{n_k-1}{2n_k} +
 \frac{1}{2}\biggl( m_P(g_t)^2- \sum_{k=1}^r m_P (f_k)^2 \biggr)\biggr]\\
&=&\frac{1}{2} \sum_{P \in Sing(g_t)}\biggl[
  m_P(g_t)^2 -  \sum_{k=1}^r  \frac{m_P(f_k)^2}{n_k}\label{bnm}\biggr].
\end{eqnarray}

On the other hand,
the right-hand side of Bezout's Theorem (equation (\ref{bezoutseveral})) is
\begin{equation}\label{bez6}
\sum_{k=1}^r \sum_{1\leq i<j\leq n_k} \deg(f_{k,i})\deg(f_{k,j})+
\sum_{1\leq k<l\leq r} \sum_{\substack{1\leq i\leq n_k\\1\leq j\leq n_l}}\deg(f_{k,i})\deg(f_{l,j}).
\end{equation}
Since each $f_{k,i}$ has the same degree for all $i$,
the first term is equal to
\[
\sum_{k=1}^r  \deg(f_k)^2\ \frac{n_k-1}{2n_k}=
\frac{1}{2} \sum_{k=1}^r  \deg(f_k)^2 - \frac{1}{2} \sum_{k=1}^r  \frac{\deg(f_k)^2}{n_k}.
 \]
 Note that
\begin{eqnarray*}
(\deg (g_t))^2&=&\biggl( \sum_{k=1}^r  \deg (f_k) \biggr)^2\\
&=&  \sum_{k=1}^r \deg (f_k)^2 +2 \biggl( \sum_{1\leq k<l\leq r} \deg (f_k) \deg (f_l) \biggr)\\
&=&   \sum_{k=1}^r \deg (f_k)^2 +2 \sum_{1\leq k<l\leq r} \biggl( \sum_{i=1}^{n_k} \deg (f_{k,i})\biggr)
  \biggl( \sum_{j=1}^{n_l} \deg (f_{l,j})\biggr)\\
  &=&  \sum_{k=1}^r \deg (f_k)^2 +2
  \sum_{1\leq k<l\leq r} \sum_{\substack{1\leq i\leq n_k\\1\leq j\leq n_l}}\deg(f_{k,i})\deg(f_{l,j}).
\end{eqnarray*}
Substituting both of these into (\ref{bez6}) shows that (\ref{bez6}) is equal to
\begin{equation}\label{otherside}
\frac{1}{2}\biggl(\deg(g_t)^2-\sum_{k=1}^r  \frac{\deg(f_k)^2}{n_k} \biggr).
\end{equation}
 Comparing (\ref{otherside}) and (\ref{bnm}),
 so far we have shown that Bezout's Theorem implies the following inequality:
 \[
 \deg(g_t)^2-\sum_{k=1}^r  \frac{\deg(f_k)^2}{n_k} \leq
 \sum_{P \in Sing(g_t)}\biggl[
  m_P(g_t)^2 -  \sum_{k=1}^r  \frac{m_P(f_k)^2}{n_k}\biggr].
 \]
 Finally, using (\ref{eq10})
and  Lemma \ref{SquareOfDegrees&SquareOfMultiplicities} to compare both sides
term by term,
this is a contradiction.
$\square$\bigskip

\section{A Counterexample}

We have found with MAGMA \cite{BCP}
that $t=205$ is a counterexample to Conjecture 3.
In this case, $g_t (x,y)$ factors into two factors over $\mathbb{F}_2$.
One of the factors is

\medskip

$x^{10} + x^9y + x^9 + x^8y^2 + x^8y + x^8 + x^6y^4 + x^6y^3 + x^6y^2 +
       x^6y + x^6 + x^5y^5 + x^5 + x^4y^6 + x^4y^4 + x^4y^3 + x^4y^2 +
       x^4 + x^3y^6 + x^3y^4 + x^3y^3 + x^3y + x^2y^8 + x^2y^6 + x^2y^4
       + x^2y^2 + x^2 + xy^9 + xy^8 + xy^6 + xy^3 + xy + x + y^{10} + y^9 +
       y^8 + y^6 + y^5 + y^4 + y^2 + y + 1$

\medskip

\noindent and the other factor has many more terms!
Note in this case that $i=2$ and $\ell=51$, so $gcd(\ell, 2^i-1)=3$.
In this case we have $1<gcd(\ell, 2^i-1)<\ell$.

{}
\end{document}